\documentclass[%
 oneclum,
 amsmath,amssymb,
]{revtex4-1}

\usepackage{graphicx}
\usepackage{dcolumn}
\usepackage{bm}

\usepackage[utf8]{inputenc}
\usepackage[T1]{fontenc}
\usepackage{mathptmx}
\usepackage{etoolbox}
\usepackage{color}
\usepackage{CJK}
\usepackage{comment}
\makeatletter
\def\@email#1#2{%
 \endgroup
 \patchcmd{\titleblock@produce}
  {\frontmatter@RRAPformat}
  {\frontmatter@RRAPformat{\produce@RRAP{*#1\href{mailto:#2}{#2}}}\frontmatter@RRAPformat}
  {}{}
}%
\makeatother

\usepackage{url}

\DeclareMathOperator{\diag}{diag}

\begin{document}

\preprint{AIP/123-QED}

\title[Development of Advanced Photon Calibrator for Kamioka Gravitational wave detector (KAGRA]{Development of Advanced Photon Calibrator for Kamioka Gravitational wave detector (KAGRA)}
\author{Y. Inoue}
\email{iyuki@ncu.edu.tw}
 \affiliation{ 
Physics department, National Central University, Taoyuan 32001, Taiwan.
}%
\affiliation{%
Center for High Energy and High Field Physics(CHiP),  National Central University, Taoyuan 32001, Taiwan.
}%
\affiliation{%
Institute of Physics, Academia Sinica, Taipei 11529, Taiwan.
}%

\affiliation{%
High Energy Accelerator Research Organization (KEK), Ibaraki 305-0801, Japan.
}%

\author{B.H. Hsieh}%
\affiliation{ 
Institute for Cosmic Ray Research, The University of Tokyo, Chiba 277-8582, Japan.
}%
\author{K.H. Chen}
\affiliation{ 
Physics department, National Central University, Taoyuan 32001, Taiwan.
}%
\affiliation{%
Center for High Energy and High Field Physics(CHiP),  National Central University, Taoyuan 32001, Taiwan.
}%

\affiliation{%
Molecular Sciences and Technology, Taiwan International Graduate Program, Academia Sinica, National Central University, Taipei, Taiwan.
}%
\affiliation{%
Institute of Atomic and Molecular Science, Academia Sinica, Taipei, 10617, Taiwan.
}%

\author{Y.K. Chu}%
\affiliation{%
Institute of Physics, Academia Sinica, Taipei 11529, Taiwan.
}%
\author{K. Ito}%
\affiliation{%
Department of Physics, University of Toyama, Toyama 930-8555, Japan
}%
\author{C. Kozakai}%
\affiliation{%
High Energy Accelerator Research Organization (KEK), Ibaraki 305-0801, Japan.
}%
\author{T. Shishido}%
\affiliation{%
SOKENDAI (The Graduate University for Advanced Studies), Kanagawa 240-0115, Japan.
}%
\author{Y. Tomigami}%
\affiliation{%
Department of Physics, Graduate School of Science, Osaka City University, Osaka 558-8585,~Japan.
}%
\author{T. Akutsu}%
\affiliation{%
National Astronomical Observatory of Japan (NAOJ), 181-8588, Tokyo, Japan.
}%
\author{S. Haino}%
\affiliation{%
Institute of Physics, Academia Sinica, Taipei 11529, Taiwan.
}%
\author{K. Izumi}%
\affiliation{%
JAXA Institute of Space and Astronautical Science, Chuo-ku, Sagamihara City, Kanagawa 252-0222, Japan.
}%
\author{T. Kajita}%
\affiliation{ 
Institute for Cosmic Ray Research, The University of Tokyo, Chiba 277-8582, Japan.
}%
\author{N. Kanda}%
\affiliation{%
Department of Physics, Graduate School of Science, 
 Osaka Metropolitan University, Osaka 558-8585}
\affiliation{%
Nambu Yoichiro Institute of Theoretical and Experimental Physics (NITEP), Osaka Metropolitan University, Osaka 558-8585, Japan}
\author{C.S. Lin}%
\affiliation{%
Institute of Physics, Academia Sinica, Taipei 11529, Taiwan.
}%
\author{F.K. Lin}%
\affiliation{%
Institute of Physics, Academia Sinica, Taipei 11529, Taiwan.
}%
\author{Y. Moriwaki}%
\affiliation{%
Department of Physics, University of Toyama, Toyama 930-8555, Japan
}%
\author{W. Ogaki}%
\affiliation{ 
Institute for Cosmic Ray Research, The University of Tokyo, Chiba 277-8582, Japan.
}%
\author{H.F.~Pang}%
 \affiliation{ 
Physics department, National Central University, Taoyuan 32001, Taiwan.
}%
\affiliation{%
Center for High Energy and High Field Physics(CHiP),  National Central University, Taoyuan 32001, Taiwan.
}%

\author{T. Sawada}%
\affiliation{%
Nambu Yoichiro Institute of Theoretical and Experimental Physics (NITEP), Osaka Metropolitan University, Osaka 558-8585, Japan}
\author{T. Tomaru}%
\affiliation{%
National Astronomical Observatory of Japan (NAOJ), 181-8588, Tokyo, Japan.
}%
\affiliation{ 
Institute for Cosmic Ray Research, The University of Tokyo, Chiba 277-8582, Japan.
}%
\affiliation{%
SOKENDAI (The Graduate University for Advanced Studies), Kanagawa 240-0115, Japan.
}%
\affiliation{%
High Energy Accelerator Research Organization (KEK), Ibaraki 305-0801, Japan.
}%
\author{T. Suzuki}%
\affiliation{ 
Institute for Cosmic Ray Research, The University of Tokyo, Chiba 277-8582, Japan.
}%
\author{S. Tsuchida}%
\affiliation{Department of Physics, Graduate School of Science, 
 Osaka Metropolitan University, Osaka 558-8585}%
\author{T. Ushiba}%
\affiliation{ 
Institute for Cosmic Ray Research, The University of Tokyo, Chiba 277-8582, Japan.
}%
\author{T. Washimi}%
\affiliation{%
High Energy Accelerator Research Organization (KEK), Ibaraki 305-0801, Japan.
}%
\author{T.~Yamamoto}%
\affiliation{ 
Institute for Cosmic Ray Research, The University of Tokyo, Chiba 277-8582, Japan.
}%
\author{T. Yokozawa}%
\affiliation{ 
Institute for Cosmic Ray Research, The University of Tokyo, Chiba 277-8582, Japan.
}%
\date{\today}

\begin{abstract}
The Kamioka Gravitational wave detector (KAGRA) cryogenic gravitational-wave observatory has commenced joint observations with the worldwide gravitational wave detector network. Precise calibration of the detector response is essential for accurately estimating the parameters of gravitational wave sources. The photon calibrator is a crucial calibration tool used in Laser Interferometer Gravitational-wave Observatory(LIGO), Virgo, and KAGRA, and it was utilized in the joint observation 3 with GEO600 in Germany in April 2020. In this paper, KAGRA implemented three key enhancements: a high-power laser, a power stabilization system, and remote beam position control. KAGRA employs a 20~W laser divided into two beams that are injected onto the mirror surface. By utilizing a high-power laser, the response of the detector at kHz frequencies can be calibrated. To independently control the power of each laser beam, an optical follower servo was installed for power stabilization. The optical path of the photon calibrator's beam positions was controlled using pico-motors, allowing for the characterization of the detector's rotation response. Additionally, a telephoto camera and quadrant photodetectors were installed to monitor beam positions, and beam position control was implemented to optimize the mirror response. In this paper, we discuss the statistical errors associated with the measurement of relative power noise. We also address systematic errors related to the power calibration model of the photon calibrator and the simulation of elastic deformation effects using finite element analysis. Ultimately, we have successfully reduced the total systematic error from the photon calibrator to 2.0~\%.
\end{abstract}

\maketitle

\section{Introduction}
The detection of gravitational waves (GWs) has the potential to lay the foundation for future advancements in our understanding of physics, including general relativity, nuclear physics, cosmology, and astrophysics~\cite{PhysRevLett.116.061102}. Calibrated gravitational waveforms with low statistical and systematic errors are particularly important as they can provide us with significant insights into new physics. The Advanced LIGO~\cite{0264-9381-32-7-074001}  and Advanced Virgo~\cite{0264-9381-32-2-024001} projects have successfully measured gravitational waves that are consistent with numerical relativity simulations. Furthermore, the international joint observation involving Advanced LIGO, Advanced Virgo, Kamioka Gravitational wave detector (KAGRA)~\cite{0264-9381-29-12-124007, PhysRevD.88.043007, 10.1093/ptep/ptaa125} , and LIGO India is expected to provide crucial information about gravitational wave sources, including their masses, spins, localization, redshift, and polarizations.

KAGRA is a 3-km Large-Scale Cryogenic Gravitational Wave Telescope located in Kamioka, Gifu prefecture, Japan~\cite{0264-9381-29-12-124007, PhysRevD.88.043007, 10.1093/ptep/ptaa125}. It employs two unique approaches: an underground site and cryogenic mirrors. The stable underground environment and cryogenic mirrors contribute to reducing seismic noise and thermal noise, respectively.

GWs cause differential variations in the arm length of the telescope, resulting in power modulations in the detector readout. The power fluctuations measured by a photodetector serve as the GW readout signal and also as an error signal for controlling the differential arm length. To ensure stable operation of the instrument, feedback control of the differential arm length is required. This control is achieved by digitizing the readout signal, applying a set of digital filters, and sending the opposite phase signal of the filtered signal as a control signal to the test mass actuators. Therefore, in order to accurately estimate the equivalent GW strain sensed by the interferometer, characterization and correction with calibration for the feedback control loop are necessary~\cite{PhysRevD.88.043007, 0264-9381-34-22-225001, 10.1093/ptep/ptab018, Sun_2020}. By modeling the differential arm length using parameters obtained from calibration, the calibrated gravitational wave strain can be reconstructed~\cite{Sun_2020}.

Calibration uncertainties directly contribute to errors in the absolute GW signal. The most significant impact of calibration uncertainties is observed in determining the distance to the GW source. Since the estimation of the GW source population relies on the cube of the source distance, calibration uncertainties also introduce uncertainties in population estimations~\cite{Abbott:2017xzu, Schutz_1986, Feeney:2018mkj}. Additionally, calibration uncertainties affect coordinate reconstruction, especially when only three detectors in the global GW detector network detect the GW signal. This is often the case due to the directional dependence of interferometer sensitivity. The impact of calibration uncertainties is more prominent in high signal-to-noise ratio events, where the angular resolution is less affected by detector noise~\cite{Abbott2016,0264-9381-34-22-225001, Estevez_2018}.
To address calibration, the photon calibrator (Pcal) is utilized in LIGO, Virgo, and KAGRA to calibrate the interferometer response. 

\begin{figure}[h!t]
\centering
\includegraphics[width=.8\linewidth]{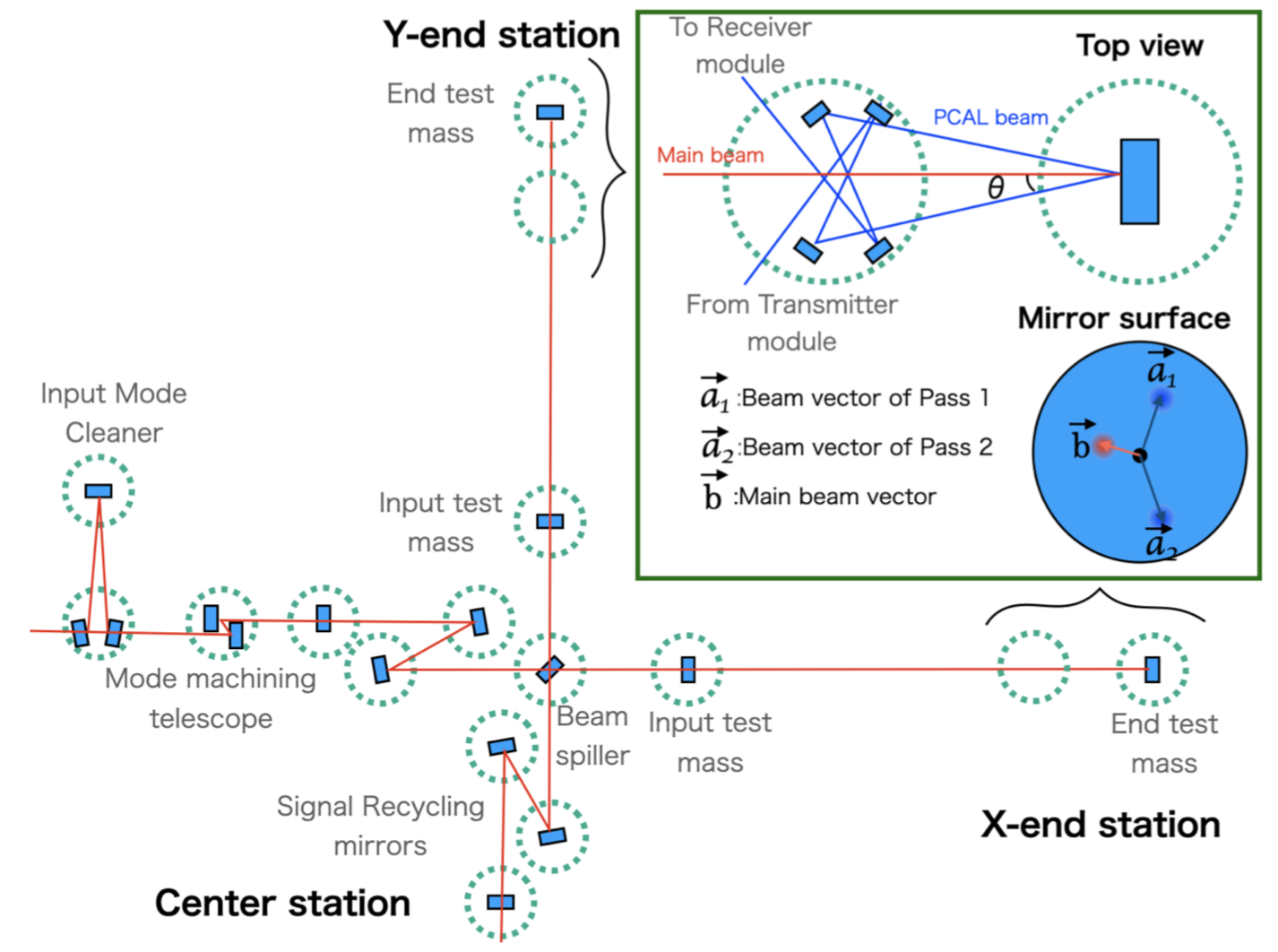}
\caption{Schematic view of KAGRA interferometer. Photon calibrators are placed at X-end and Y-end stations. Top right view shows the schematic view of photon calibrator and end test mass.}
\label{fig0}
\end{figure}

The Pcal induces surface modulation of the mirror through photon pressure. In the joint observation run 3involving KAGRA and GEO600 (O3GK) in April 2020~\cite{10.1093/ptep/ptaa120, 10.1093/ptep/ptaa125, 10.1093/ptep/ptab018, 10.1093/ptep/ptx180}, the photon calibrator served as the primary calibrator for KAGRA. The details of the calibration overview during O3GK are provided in the summary paper~\cite{10.1093/ptep/ptab018}, while the initial characterization of photon calibrator instruments is described elsewhere~\cite{bin-hua, cory}. The Pcal achieves mirror displacement by injecting a power-stabilized laser with intensity modulation. The main applications of the photon calibrator are (i) calibration of the interferometer response, (ii) monitoring time-dependent changes in the interferometer response, and (iii) hardware injection to verify the analysis pipeline~\cite{PhysRevD.95.062002,cory}. The displacement of the mirror can be expressed as follows: 
\begin{align}
\label{eq1}
X = \frac{2P \cos \theta }{c} S_{\text{tot}} (f, \vec{a}, \vec{b}), 
\end{align}
where $P$ is the absolute laser power, $\theta $ is the incident angle of the Pcal laser, and $c $ is the speed of light~\cite{doi:10.1063/1.4967303,0264-9381-27-8-084024,0264-9381-26-24-245011}. $S_{\text{tot}} (f, \vec{a}, \vec{b})$ is the force-to-displacement transfer function of the suspended pendulum. The complete form can be defined as
\begin{align}
\label{eq2}
S_{\text{tot}} (f, \vec{a}, \vec{b})=S_\text{Len} (f)+S_\text{Rot} (f, \vec{a}, \vec{b})+S_\text{Ela} (f, \vec{a}, \vec{b}),
\end{align}
where $S_\text{Len} (f)$, $S_\text{Rot} (f, \vec{a}, \vec{b})$ and $S_\text{Ela} (f, \vec{a}, \vec{b})$  are the transfer functions of pendulums for displacement, rotation and elastic deformations of mirror, respectively. In our observation frequency, we assumed that the pendulum transfer function was free mass motion. The transfer function is obtained from the following equations:
\begin{align}
\label{eq3}
S_\text{Len}(f)=1/(M\omega^2)
\end{align}
and
\begin{align}
S_\text{Rot}(f)=\frac{\vec{a} \cdot \vec{b}}{I \omega^2} \label{eqrot}
\end{align}
where $M $ and $I =   Mh^{2} /12  +   Mr^{2}/4$ are the mass and moment of inertia of the test mass, $h $ and $r $ are the thickness and radius of the test mass, $\omega $
is the angular frequency of the laser power modulation, and $\vec{a}$ and $\vec{b}$ are the vectors pointing to the center of force of the Pcal beams and the main interferometer beam, respectively~\cite{doi:10.1063/1.4967303,0264-9381-27-8-084024,0264-9381-26-24-245011}. Figure~\ref{fig0} shows the schematic view of KAGRA interferometer and definition of beam vectors. The definition of beam position vectors are shown in Fig.~\ref{fig1}. The transfer function with elastic deformation is described below.~\cite{Hild_2007}
\begin{align}
S_\text{Ela}(f, \vec{a}, \vec{b}) =\frac{\iint d\xi d\eta G(\xi, \eta, \vec{b}) D(\xi, \eta, \vec{a};f)} {\iint d \xi d \eta G(\xi, \eta, \vec{b})}
\end{align}
where $D(\xi, \eta, \vec{a};f)$ is the deformation of test mass surface normalized by the injected power, and $(\xi, \eta)$ are coordinate parameters, $G(\xi, \eta, \vec{b})$ is Gaussian beam profile of main laser beam.

The response to excitation forces can be represented by an appropriate linear combination of normal modes. In particular, the butterfly mode and drumhead mode are the main contributors to elastic deformation. However, these symmetric deformation effects can be mitigated by applying at least two beams that are diametrically opposed and sufficiently displaced from the center of the test mass. This technique eliminates $S_\text{Rot}$ and $S_\text{Eal}$, leaving only $S_\text{Len}$, which is necessary for calibration. This scheme was tested and implemented in LIGO with an Advanced LIGO photon calibrator~\cite{doi:10.1063/1.4967303}.

We monitored the time dependency of the transfer function using the calibration pipelines~\cite{10.1093/ptep/ptab018}. This method is also useful for monitoring the time dependency of other response functions, such as actuation response, optical gain, and cavity pole frequency. By utilizing this information, we can estimate the uncertainty of gravitational wave strain since the relative response function is proportional to the relative gravitational wave strain.

The displacement of the mirror is proportional to the absolute laser power. Therefore, we needed to calibrate the absolute power of the Pcal laser. LIGO's power standard is calibrated by NIST every year, with a relative uncertainty of 0.32~\%. On the other hand, NIST compares the absolute laser power response with several countries, resulting in an absolute uncertainty of about 3~\%~\cite{EUROMET, Bhattacharjee_2020}. The power-stabilized laser used in the Pcal also provided information for consistency checks of the response function. By using the Pcal read-back signal as the response of the receiver module output, we were able to estimate the expected response of the end test mass. The definition of the PCAL read-back signal is explained in Sec.~\ref{model}. We can compare the expected $h(t)$ obtained from the read-back signal with the estimated $h(t)$ from the reconstruction pipeline of KAGRA.

Recently, the gravity field calibrator and Newtonian Calibrator methods have been proposed for absolute calibration ~\cite{PhysRevD.98.022005, Estevez_2021, Estevez_2018} . The gravity field calibrator consists of a rotating disk with quadrupole mass distribution, which creates a changing gravity gradient around the end test mass. The demonstration of this method will be carried out in a future gravitational wave experiment.

In this paper, we summarized the photon calibrator instruments of KAGRA, including the measurement of relative intensity noise and relative harmonic noise. Particularly, the characterization of Pcal with a 20 W laser is the first demonstration in the research field. In Section 2, we explained the specifications and design of the system. In Section 3, we discussed the measurement results of relative power noise and harmonic noise. In Section 4, we obtained the systematic errors.

\section{INSTRUMENTS}
\begin{figure}[h!t]
\centering
\includegraphics[width=.8\linewidth]{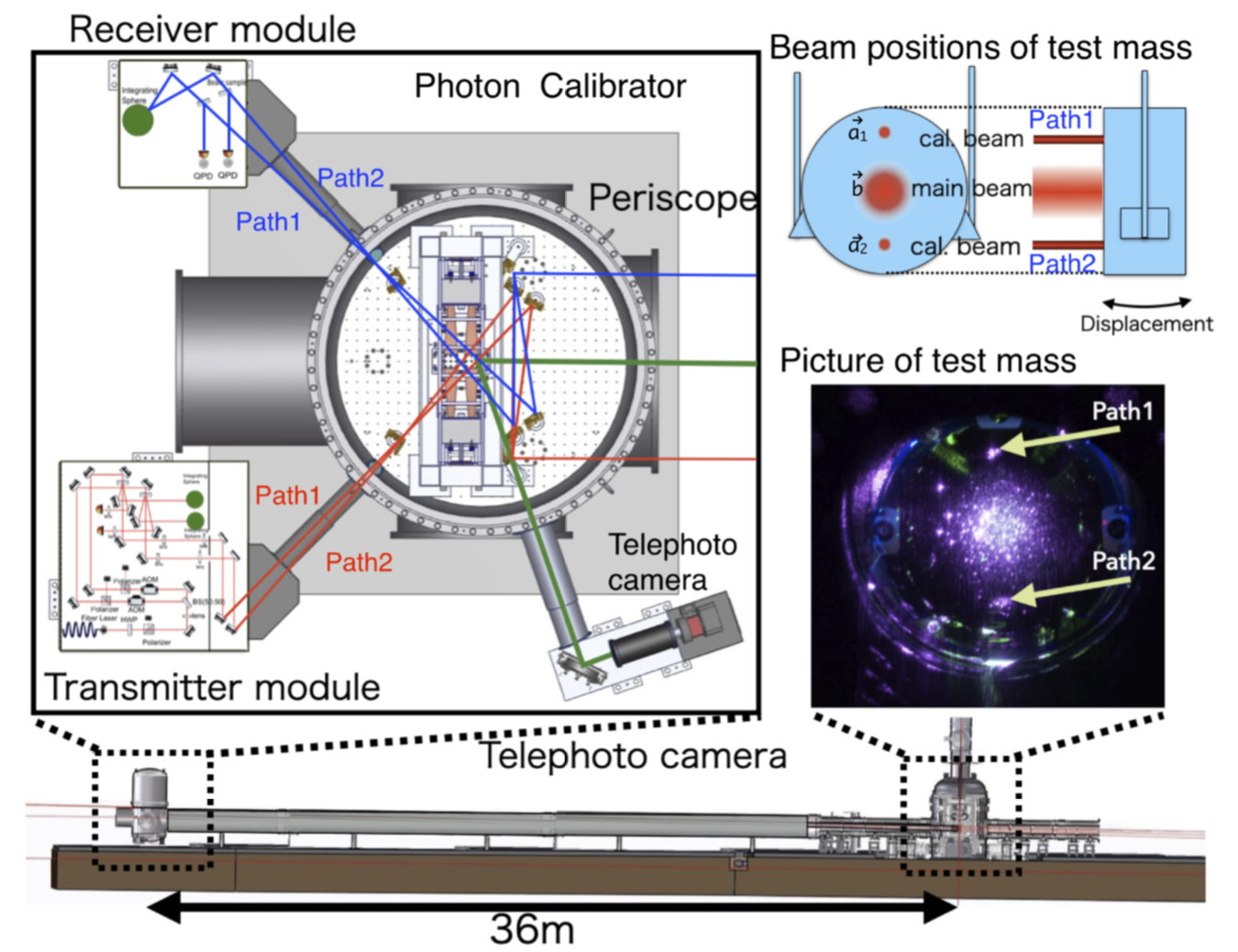}
\caption{Schematic view of the KAGRA photon calibrator. Right upper figure shows the definition of beam positions. $\vec{a}_1$ and $\vec{a}_2$ are beam position of path 1 and path 2, respectively. $\vec{b}$ is beam position of main beam. The origin of these vector is defined on the center of mirror surface.}
\label{fig1}
\end{figure}

\begin{figure}[h!t]
\centering
\includegraphics[width=.8\linewidth]{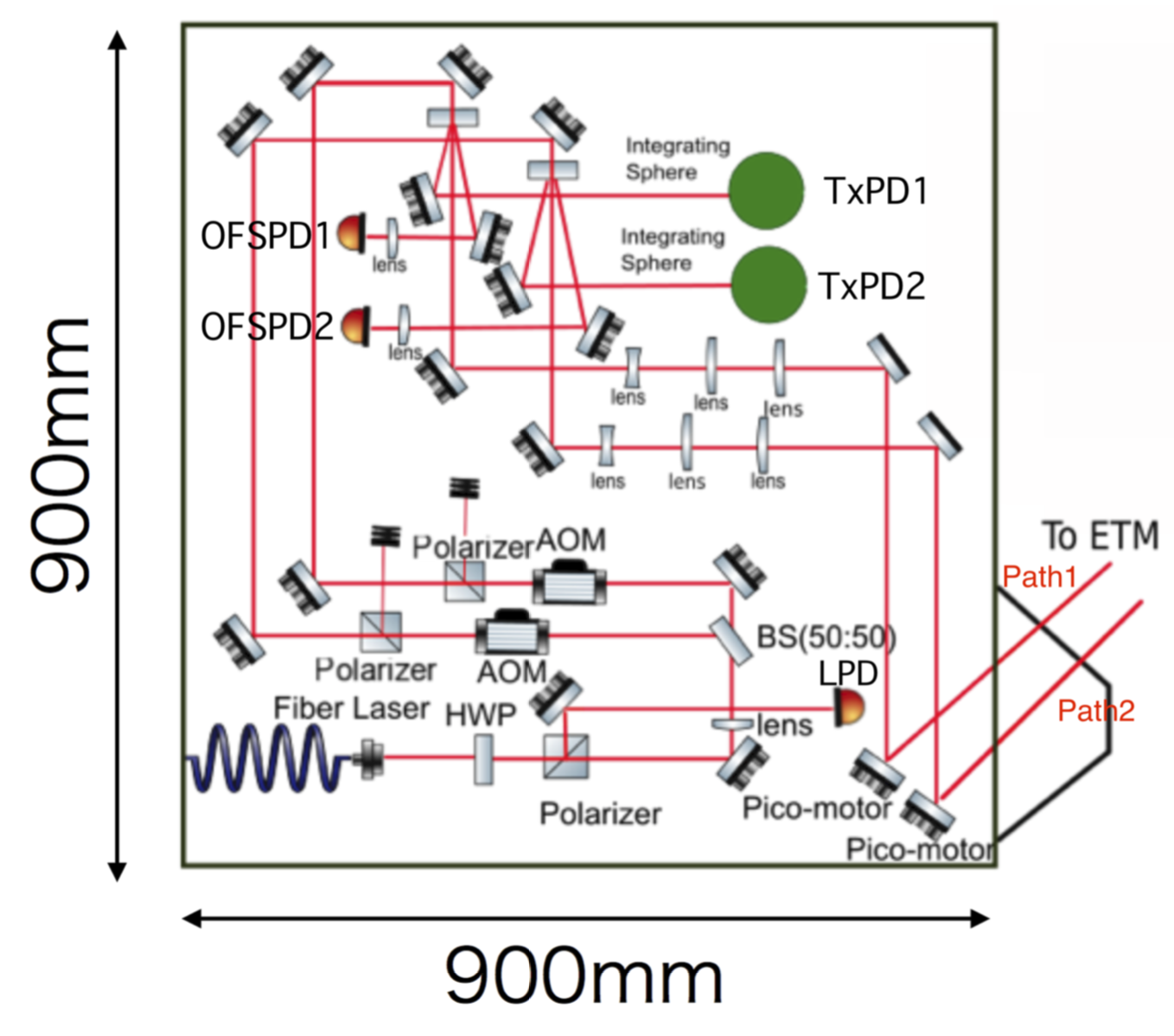}
\caption{Optical layout of the Transmitter module. We installed the CW laser on the optical table and its stabilization system. The laser beam is split by the beam splitter. The split beams are modulated with AOMs. The schematic view of feedback loop is shown in Fig.~\ref{fig6}}
\label{fig2}
\end{figure}

The KAGRA photon calibrator was positioned at the X and Y end stations, which were located 3 km from the beam splitter of the interferometer as illustrated in Fig.~\ref{fig0}. The X-end system was used for interferometer calibration, while the Y-end system was employed for hardware injection during observation~\cite{PhysRevD.95.062002}. These systems were installed at a distance of 34.957 m from the end test mass (ETM) as shown in Fig.~\ref{fig1}. The ETM is composed of Sapphire crystal. Two laser beams, referred to as Path-1 and Path-2, were injected into the ETM. The layout of the KAGRA photon calibrator is shown in Fig. ~\ref{fig1}. The photon calibrator consisted of a transmitter module (Tx module), a receiver module (Rx module), a periscope, and a telephoto camera module (TCam module). These components provided intensity-modulated laser beams, monitored the intensity, adjusted the height of the optical axis, and observed the beam spot position on the mirror surface, respectively. We utilized a continuous wave (CW) fiber laser with a maximum power of 20 W and a wavelength of 1047 nm to avoid coupling with the main laser. To stabilize and modulate the laser power, an optical follower servo (OFS) was installed in the Tx module. The beams were split in the Tx module to minimize elastic deformation caused by pushing the ETM, as shown in Fig.~\ref{fig2}. The position of each beam was measured with a telephoto camera (TCam). A summary of the specifications for the KAGRA photon calibrator is presented in Table~I.

\begin{table}[h!]
\centering
\caption{Specification summary of the KAGRA photon calibrator.Incident angle of Pcal is defined by the interval of periscope mirror and distance between Pcal and end test mass. Beam waist of input laser is measured on the Tx module. Definition of beam position corresponds to $\vec{a}=\vec{a_1}+\vec{a_2}$ as shown in Eq.\ref{eq2} and Fig.\ref{fig1}. The origin of vectors is defined on the center of mirror surface.}
\label{pcal_over}
\begin{tabular}{cc}
\hline
Mirror material & Sapphire \\
Mirror mass &  22.95  $\pm$  0.01~kg \\
Mirror diameter & 220~mm \\
Mirror thickness & 150~mm \\
Distance from Pcal to test mass &  34.957 $\pm$  0.01~m \\
Maximum laser power & 20~W \\
Pcal laser wavelength & 1047~nm \\
Incident angle &  0.839 $\pm$ 0.023~deg \\
Beam waist of input laser &  265.57 $\pm$ 0.04~[um]\\
Beam position $\vec{a_1}$ (Top) & (0~mm, 76~mm) \\
Beam position $\vec{a_2}$ (bottom) & (0~mm, $-$76~mm)\\
\hline
\end{tabular}
\end{table}

\subsection{Transmitter module}

The transmitter module served as the input optical system for modulating and stabilizing the laser power. The optical components of the transmitter module were mounted on a 900~mm  $\times $  900~mm optical table, as depicted in Fig. ~\ref{fig2}. To mitigate atmospheric fluctuations, the system was enclosed with aluminum plates coated with Tobika black to absorb scattered light. We utilized a 20 W Yb continuous wave (CW) laser manufactured by Keopsys, whose model number is CYFL-TERA-20-LP-1047-AM1-RGO-OM1-T305-C1. The laser had a typical beam waist of 265.6 $\pm $ 0.04 $\mu$m and a beam quality factor $M^2$ of 1.06 $\pm $ 0.01. The laser beams were split using a beam splitter and then modulated by ISOMET Acousto-Optic modulators (AOM), whose part number is M1080-T80L-M. With the AOM, we were able to control the laser power for each path from 0 W to 10 W. We adjusted the offset at half of the maximum AOM response. Photodiodes were used to monitor the power of the beams. We employed InGaAs PIN photodiodes with a diameter of 3.0 mm, manufactured by Excelitas. The part number of photo detector is C30665GH. In the transmitter module, five photodiodes were mounted with a trans-impedance amplifier unit: OFSPD1, OFSPD2, TxPD1, TxPD2, and LPD, as shown in Fig.~\ref{fig2}. OFSPD1 and OFSPD2 were utilized to obtain the feedback signal connected to the AOM for laser power stabilization. The sensed laser powers by OFSPD1 and OFSPD2 were attenuated to 1 mW. The absolute shot noise of the sensor can be described as  $\sqrt{2P\eta e}$, where $P$ is the detection power, $\eta$ is the responsivity of the photodiode, and $e$ is the elementary electric charge. Shot noise is characterized by the number of photons, and increasing the power can decrease the shot noise. However, the noise level of the laser does not meet the shot noise level. In an ideal case, the final sensitivity of the laser with the stabilization system is limited by shot noise. The dominant noise of this system is explained in Section~III. The signals detected by OFSPD1 and OFSPD2 were fed to the optical follower servo. LPD served as a photo detector for monitoring laser noise as an out-of-loop signal. TxPD1 and TxPD2 were employed to monitor absolute power, and their responses were calibrated using NIST standard laser. The beams were collimated by three lenses to reduce spherical aberration, and the beam radius on the ETM was 3.5~mm. Pico-motors were installed to remotely control the beam position on the mirror.

\subsection{Periscope}
To precisely control the beam position on the mirror surface, we incorporated a periscope within the vacuum chamber. A 100 mm fused silica view window was mounted on the vacuum chamber to allow observation. The periscope consisted of a structure that held the mirrors and a baffle to eliminate scattered light. We mounted 12 mirrors with a diameter of 3~inches inside the chamber. Each mirror had a reflectance of 99.95\% at 1047 nm and was manufactured by Opto-Sigma. The incident angle of the beam was set to 0.839 degrees, and we also installed a periscope for the telephoto camera on the same structure.

\begin{figure}[h!t]
\centering
\includegraphics[width=.75\linewidth]{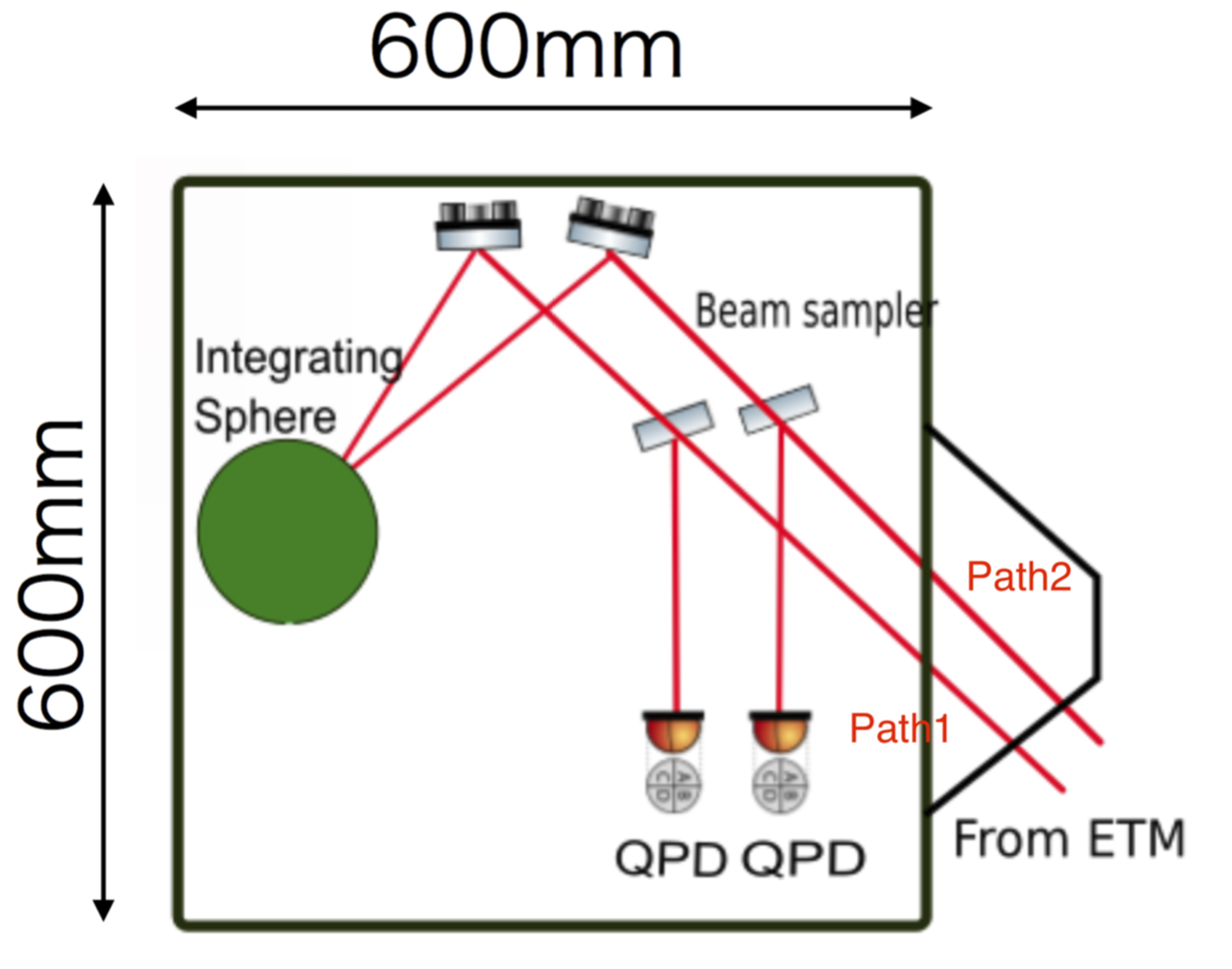}
\caption{Optical layout of receiver module. The integrating sphere and QPDs are mounted on the optical table.}
\label{fig4}
\end{figure}

\subsection{Receiver module}

The receiver module had the primary function of monitoring the reflected absolute power and beam position, as illustrated in Fig.\ref{fig4}. All the optical components of the receiver module were securely mounted on a 600mm $\times $ 600mm optical table. To facilitate this monitoring process, we integrated a 6-inch integrating sphere with a photodetector that shared the same design as the TxPD. The detailed method for absolute power calibration is outlined in SectionIV. In addition, we installed a beam sampler and utilized quadrant photodetectors (QPDs), PDQ80A, to effectively track any potential drift in beam positions. The QPDs were employed as a tilt sensing optical lever and boasted a sensitivity of 10 nrad each. By combining the functionality of the pico-motors on the transmitter module with the QPDs, we were able to both control and closely monitor the beam positions throughout the system.

\subsection{Telephoto camera}

\begin{figure}[!t]
\centering
\includegraphics[width=.8\linewidth]{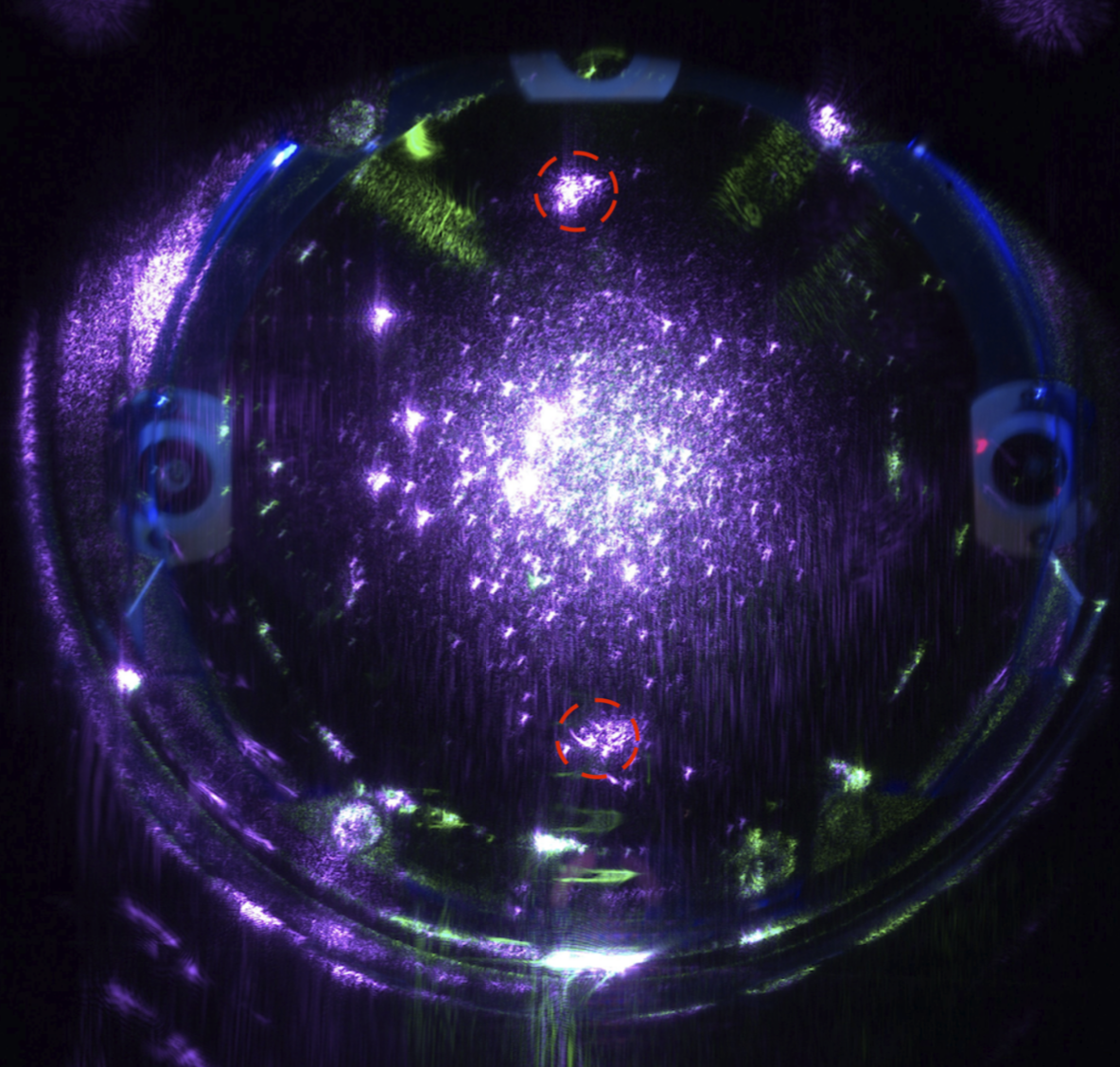}
\caption{The picture of end test mass. Upper and lower in dashed circles show the Pcal laser beams. These positions are monitored by Telephoto Camera.}
\label{fig5}
\end{figure}

To accurately measure the beam position, we utilized a camera system and image analysis. Figure~\ref{fig5} shows an example of the injected beam for the ETMX, where we injected the laser 76~mm above and below the center position. It is important to note that these positions of the laser injection were chosen to counteract the rotation and elastic deformation effects induced by the beams. To cancel out these effects, it was necessary to inject two laser beams. The drift in the beam position on the ETM surface directly corresponded to systematic errors arising from rotation and elastic deformation.

To monitor the beam position, we employed a telephoto camera (TCam), which combined a telescope with a high-resolution camera. The camera system was placed at a distance of 34.957m from the ETM. To achieve the desired accuracy, we used a Moonlight focuser to finely tune the focal point on the mirror surface. For our requirements, we selected a Nikon D810 digital camera, which features a 36-megapixel resolution and a 35.9 $\times $ 24.0~mm CMOS sensor. Each pixel in the camera image corresponded to a physical distance of 100 $\mu $m.

To ensure compatibility with the laser wavelength (1047nm), we removed the infrared (IR) filter from the commercial camera, as it is typically not sensitive to this wavelength. In our setup, we employed a Maksutov-Cassegrain telescope designed for observing the ETM surface. The primary mirror of the telescope had a diameter of 127mm, with a focal length of 1500mm and a focal ratio of $f/12$. Additionally, we installed an LED illuminator on the cryogenic stage near the ETM, which could be controlled remotely. This setup allowed us to obtain clear and precise monitoring of the mirror surface from a distance of 36m. By observing the mirror surface from a far location, we minimized the scattering effects caused by the small solid angle between the camera and the ETM.

\begin{figure}[h!t]
\centering
\includegraphics[width=\linewidth]{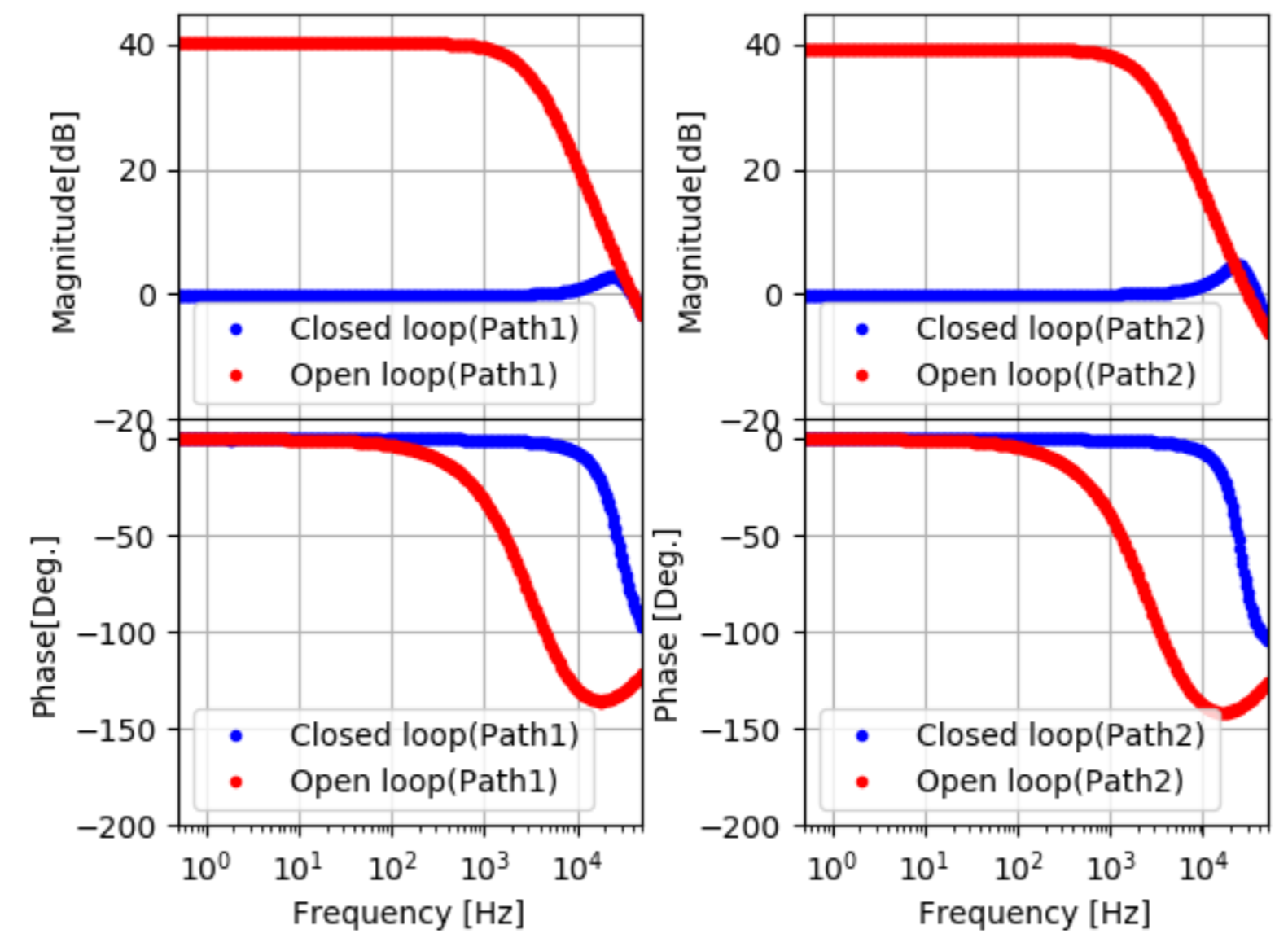}
\caption{Transfer function of Path1 and Path2. In the application of Pcal, we require the unity response and 0 degree phase differences of closed loop response in observation frequency to minimize frequency dependence of calibration signal by the control. }
\label{fig3}
\end{figure}

\section{Measurement}

The noise of the Pcal laser directly propagated to the displacement noise through the response of the suspension system. The laser exhibited a typical relative power noise of $-$120 dB/$\sqrt{\text{Hz}}$ for each path. This power noise had a more significant impact on mirror position fluctuations compared to the desired sensitivity of the KAGRA detector.

To mitigate the laser noise and reduce its impact on the system, we implemented a stabilization mechanism using servo filters, known as the optical follower servo. The transfer function of the optical follower servo is illustrated in Fig.~\ref{fig3}, with two poles set at 3 kHz and one pole at 30 kHz. The closed-loop transfer function, defined as $G/(1+G)$, where $G$ represents the open-loop transfer function, determines the overall performance of the system.

To achieve optimal performance, we configured the optical follower servo to have a unity gain frequency of 40 kHz with a 50-degree phase margin. This ensures stability and robustness in the feedback loop. Figure~\ref{fig5} shows the feedback loop diagram, where we divided the laser beams into two independent paths for individual stabilization. The detected signal at the OFSPD was fed into the servo filter.

To maximize the effectiveness of the feedback loop, it was crucial to characterize the relative power noise (RPN) and harmonic noise (HN) components. In the subsequent subsections, we provide an explanation of the measured noise sources based on the KAGRA's target sensitivity range for scientific investigations, which spans from 30 Hz to 1500 Hz.

\begin{figure}[h!t]
\centering
\includegraphics[width=.86\linewidth]{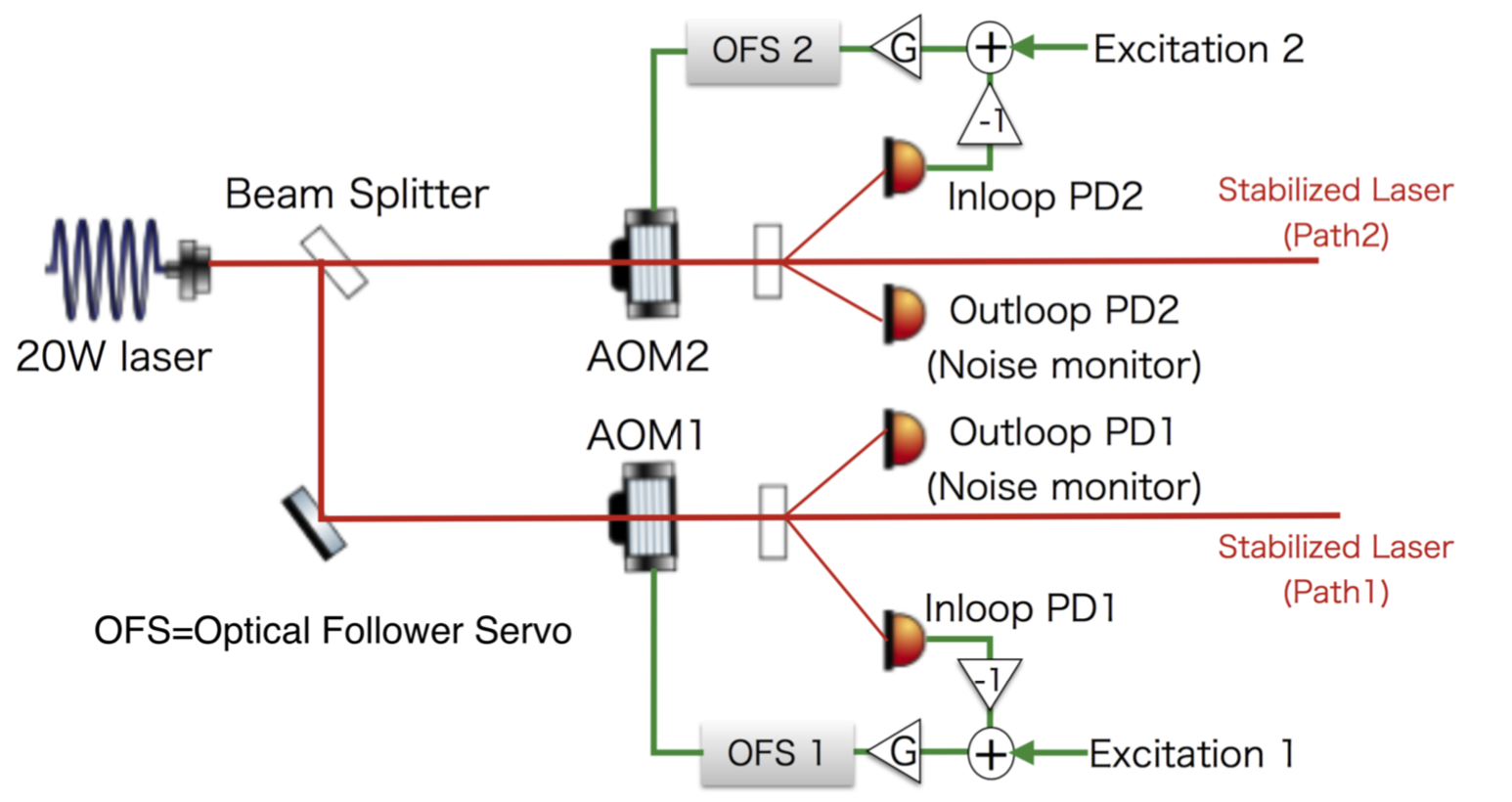}
\caption{Schematic view of the laser power stabilization system. Outloop (Inloop) PD 1 and 2 correspond to TxPD (OFSPD) 1 and 2 in Fig.~\ref{fig2}, respectively.}
\label{fig6}
\end{figure}

\begin{figure}[b!]
\centering
\includegraphics[width=\linewidth]{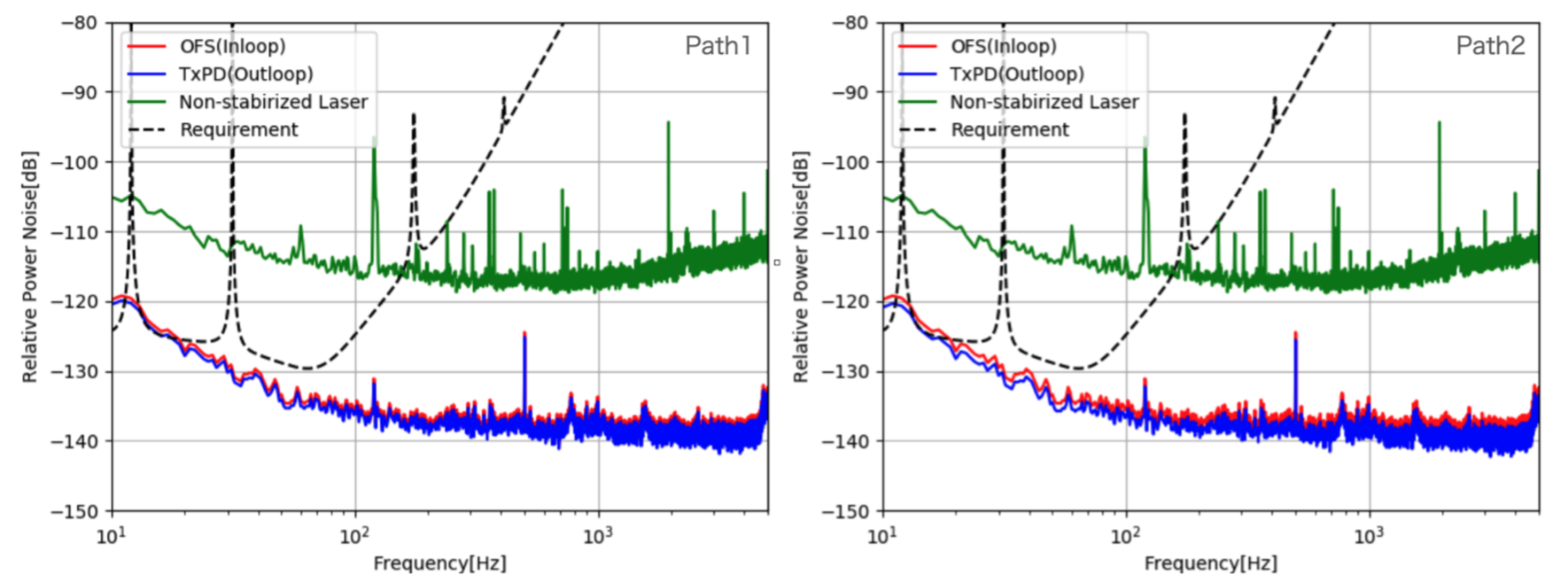}
\caption{Measured relative power noises and design sensitivity of the photon calibrator. The green curves represent the non-stabilized laser. The dashed line is the requirement from the design sensitivity of KAGRA~\cite{10.1093/ptep/ptaa125}.}
\label{fig7}
\end{figure}

\subsection{Relative power noise} \label{RPN}
The optical follower servo effectively suppressed the relative power noise (RPN) of the Pcal laser. The amplitude spectral density of each path, denoted as $G_{RPN}$, needed to satisfy the following condition:

\begin{align}
G_{RPN}\leqq \frac{Lh(f) Mc\omega^2}{10 \cos \theta P_i},
\end{align}

where $i$ represents path~1 or path~2.	
 
 $h(f)$ is the target sensitivity of KAGRA during Observation phase 3 (O3), $L = 3000$ m is the arm length of the interferometer, $M_c$ is the safety margin of 10, and $P_i$ is the power in each path. The open-loop transfer function is depicted in Fig.~\ref{fig3}.

Figure~\ref{fig7} illustrates the measured RPN for path 1 and path 2. The measured noise in the out-of-loop configuration satisfied our requirements. The dominant noise below 100 Hz corresponds to the relative power noise of the laser, which was monitored by LPD. Above 100 Hz, the noise floor was determined by the noise of a 16-bit DAC (part number PCIe-16AO16-16-FO-DF) and the electrical circuit for the offset control of the AOM. The noise floor at low frequencies was limited by the noise of the offset control circuitry.

\subsection{Harmonics noise}

During the observation, sine wave excitation was continuously injected into the calibration line. The nonlinearity of the Pcal modulation introduced higher harmonics, and their amplitudes needed to be less than 10~\% of the displacement sensitivity. However, due to the suspension system's transfer function, the amplitude of the displacement caused by Pcal higher harmonics decreased with the square of the frequency.

To meet the requirement, the ratio of the amplitude of the n-th order harmonics to that of the Pcal modulation had to be less than

\begin{align}
\frac{n^2}{1000}\frac{h(nf)}{h(f)}. \label{eq8}
\end{align}

where $h(f)$ represents the target sensitivity of KAGRA at frequency $f$. Assuming a Signal-to-Noise Ratio (SNR) of 100 for the injected Pcal modulation at frequency $f$, the product of SNR and safety margin is set to 1000.

For the $2f$ signal, the higher harmonics between 20 and 750~Hz needed to be suppressed, while for the $3f$ signal, the higher harmonics between 30 and 500 Hz had to be suppressed, based on the observation frequency range. With the optical follower servo, data were accumulated for 2 hours to suppress the noise floor.

\begin{figure}[h!t]
\centering
\includegraphics[width=\linewidth]{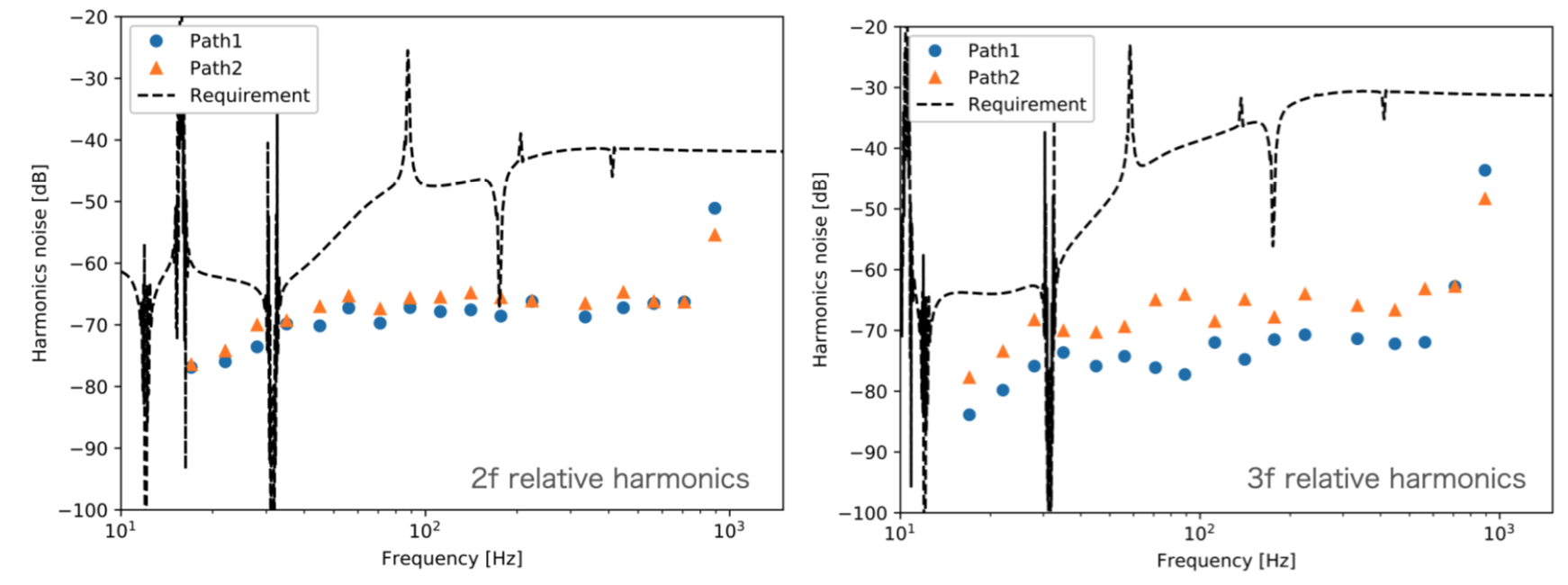}
\caption{The measured relative harmonics noises. The left and right figures show the relative harmonics of $2f$ ($n= 2$) and $3f$ ($n= 3$). The circle and triangle points correspond to Path1 and Path2. The dashed line is the requirement from the design sensitivity of KAGRA~\cite{10.1093/ptep/ptaa125} and Eq.\ref{eq9}.}
\label{fig8}\
\end{figure}

Figure~\ref{fig8} illustrates the measured relative modulation harmonics along with the requirement curve. The optical follower servo was crucial in suppressing both the power modulation at the harmonics of the Pcal fundamental frequencies and the inherent laser power fluctuations.

\section{Model} \label{model}

According to a previous study~\cite{doi:10.1063/1.4967303}, the systematic errors of the photon calibrator arise from measurement uncertainties. In order to account for these uncertainties, we incorporated the efficiency of absolute power calibration, rotation, and elastic deformation in our model, as described by Eq.~1.
To establish the parameters of the model, we needed to measure the absolute laser power and beam position. In this section, we discuss the estimation of these effects.

The photon calibrator allows us to obtain information about absolute displacement through the absolute measurement of laser power. For calibration purposes, KAGRA's photon calibrator integral spheres at the X and Y ends were compared with independent detectors that were calibrated using the NIST laser power standard ~\cite{0264-9381-26-24-245011}.
To compare the detector response of the photon calibrator with the NIST standard, we conducted the following four measurements:

\begin{itemize}
\item Absolute calibration of laser power standard (Gold Standard which belongs to LIGO) with NIST~\cite{0264-9381-26-24-245011}: $\rho_\text{GS}$ [W/V].

\item Response ratio measurement at LIGO Hanford Observatory (LHO) between Working Standard KAGRA (4 inches) and Gold Standard (4 inches) : $\alpha_{\rm WSK}$.

\item Response ratio measurement at University of Toyama between a 5.3-inch integrating sphere and Working Standard KAGRA (4~inches): $\alpha_{\text{Toyama}}$.

\item Response ratio measurement of Pcal transmitter module and receiver module with a 5.3-inch integrating sphere: $\mathcal E $.
\end{itemize}

Each displacement is function of voltage at each integrating sphere as Pcal readback signal, $\vec{V}=(V_\text{Tx1}, V_\text{Tx2}, V_\text{Rx1}, V_\text{Rx2}) $. In a usual case, we measure the sum of path-1 and path-2 response at RxPD. To determine the ratio of power at RxPD, we defines the separation efficiency, $e_{1}$ and $e_{2}$. To estimate these factors, we measure the power of path1 and path2 independently. Therefore, each term can be defined as $V_\text{Rx1} = V_\text{Rx}e_{1}$ and $V_\text{Rx2} = V_\text{Rx}e_{2}$. The displacement vector $\vec{x}=(x_\text{Tx1}, x_\text{Tx2}, x_\text{Rx1}, x_\text{Rx2}) $ is calculated as

\begin{align}
\vec{x}= \mathcal{S} \Gamma  \vec{V}, \label{eqx}
\end{align}
where $\Gamma $ is a coefficient matrix that converts voltage to force, all of which we describe in detail later. 
$\Gamma $ is the $4 \times 4$ force coefficient matrix, which is transfer matrix from output voltage of photo detectors (integral spheres) to force by photons.
The unit of components of force coefficient matrix is [N/V]. Each force coefficient can be written as
\begin{align}
\Gamma= \frac{2\cos \theta}{c} \rho
\end{align}
where $\theta $ is the incident angle of the photon calibrator as listed in Table~\ref{pcal_over}, and $c $ is the speed of light. The incident angle is determined from CAD drawings, and the variance of $\theta $ is estimated by the accuracy of the vacuum chamber.
The 1-sigma uncertainty is 0.839 $\pm $ 0.023 degree. $\rho $ is a conversion matrix from photodetectors (integral spheres) output to power of incident Pcal power, 
\begin{align}
\rho=\rho_{\rm GS} \alpha_{\text{WSK}} \alpha_\text{Toyama} \mathcal{E}\mathcal{D},
\end{align}
where $\mathcal{D}$ is the diagonal detector matrix, and each diagonal element corresponds to the calibrated detector response with the 5.3 inch integrating sphere. In the terms of $\rho$,  $\alpha_{\text{WSK}} \alpha_\text{Toyama} \mathcal{E}\mathcal{D}$ are dimensionless.Thus, the dimension of $\rho$ is [W/V]. When we included optical cross talk in the diagonal detector matrix, the off-diagonal part in the efficiency matrix ought to have been obtained. In O3GK observation, the crosstalk of each detector was negligible. Thus, we assume $\mathcal{D} = \diag(1,1,1,1)$.

The Gold Standard (GS) is a power sensor system comprised of a 4 inch integrating sphere and an InGaAs photodetector. NIST provides a summary of measurement of GS response to a 1047~nm laser. The estimated value is $\rho_{\rm GS} =$ $-$8.0985 $\pm $ 0.0259~[W/V] ~\cite{0264-9381-26-24-245011, Bhattacharjee_2020}. The calibration of GS is transferred to a 4 inch integrating sphere
with an InGaAs detector called Working Standard KAGRA (WSK). The cross-calibration setup between WSK and GS is placed in LHO. The estimated value is $ \alpha_\text{WSK}=0.2750 \pm 0.0027 $ . The measured response ratio with
University of Toyama system was
$\alpha_{\text{Toyama}} = 1.39120 \pm 0.00020 $.
$\mathcal{E}=\diag(\epsilon^{(1)}_{\rm Tx}, \epsilon^{(2)}_{\rm Tx}, \epsilon^{(1)}_{\rm Rx}, \epsilon^{(2)}_{\rm Rx})$
is the $4 \times 4$ efficiency matrix describing the loss of path 1 and path 2 for the transmitter module side and receiver module side. The measured efficiencies of efficiency matrix are described as: 
\begin{eqnarray}
\epsilon^{(1)}_{\rm Tx} = 0.980 \pm 0.040, \\
\epsilon^{(2)}_{\rm Tx} = 0.978 \pm 0.026, \\ 
\epsilon^{(1)}_{\rm Rx} = 0.984 \pm 0.040,\\
\epsilon^{(2)}_{\rm Rx} = 0.974 \pm 0.026.
\end{eqnarray}

$\mathcal{S}$ is the transfer function matrix from the force to displacement defined as
\begin{align}
\mathcal{S}=
\arraycolsep 1pt
\begin{pmatrix}
S_{\text{tot}}(f, \vec{a_1}, \vec{b}) & 0 & 0 & 0 \\
0 & S_{\text{tot}}(f, \vec{a_2}, \vec{b}) & 0 & 0 \\
0 & 0 & S_{\text{tot}}(f, \vec{a_1}, \vec{b}) & 0 \\
0 & 0 & 0 & S_{\text{tot}}(f, \vec{a_2}, \vec{b}) \\
\end{pmatrix}.
\end{align}

The analysis of elastic deformation, denoted as $S_\text{Ela}(f, \vec{a}, \vec{b})$, revealed that the calibration forces exerted by the photon calibrator beam caused local elastic deformations that significantly affected the measured displacement of the interferometer. Even materials with high stiffness, such as fused silica or sapphire, exhibited small deformations when subjected to the forces of the photon calibrator. To study this effect, modal analysis and simulation of the elastic deformation were performed using finite element analysis (FEA) software packages such as ANSYS and COMSOL.

In our analysis, we took into account the realistic structure of the KAGRA End Test Mass (ETM), which is not a perfect cylinder but has two flat cuts on both sides and two ears used for suspension~\cite{Ushiba_2021}. Based on the CAD drawing, we conducted FEA simulations on the actual KAGRA ETM structure. Two laser beams were reflected by the mirror, resulting in two beam spots located above and below the center of the mirror surface. It was found that the elastic deformation effect was minimized when the distance between the beam spots and the center was 76 mm.

\begin{figure}[h!t]
\centering
\includegraphics[width=.8\linewidth]{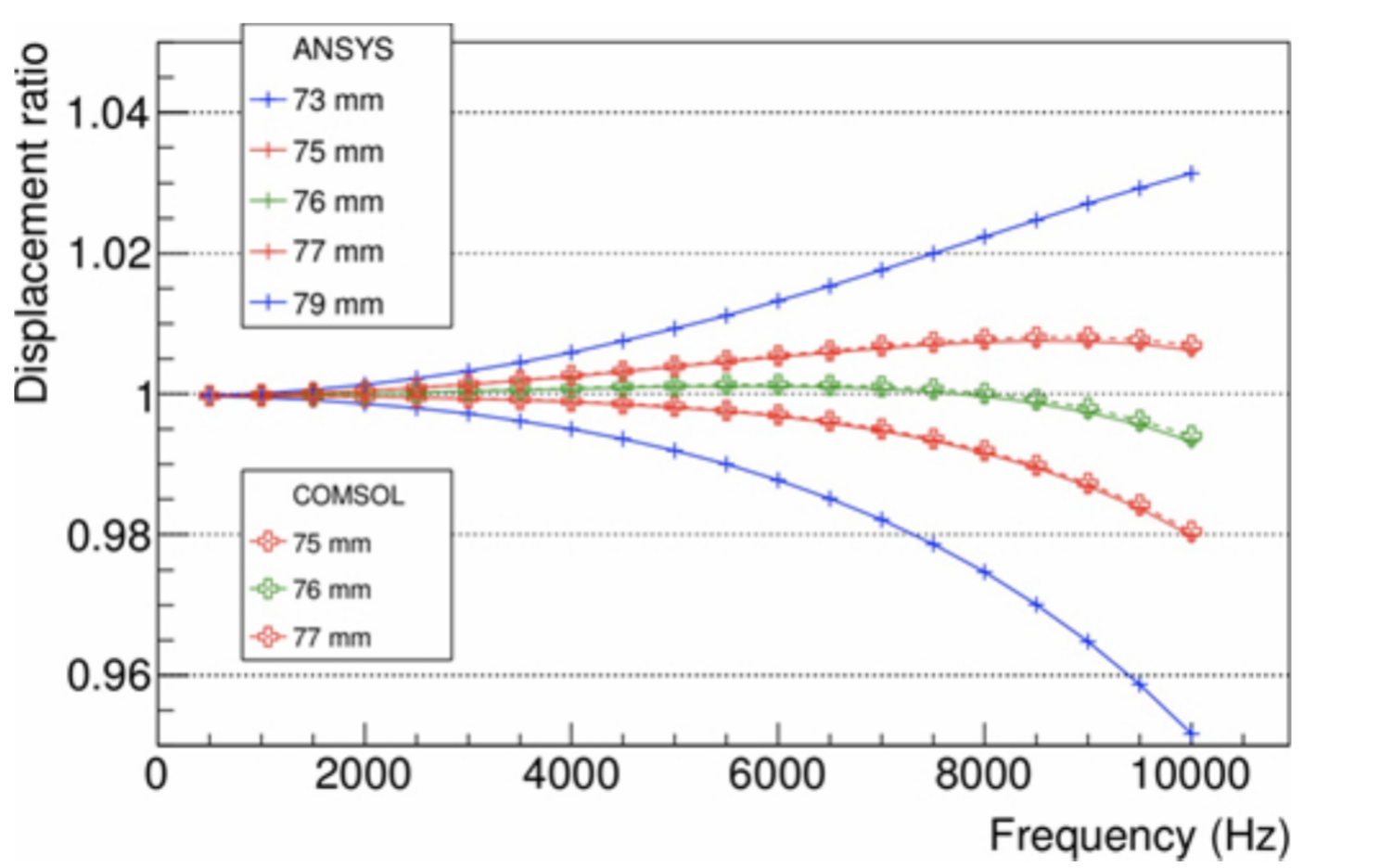}
\caption{The ratio between the total sensed motion and rigid body motion as a function of frequency for optimally positioned beams and $\pm $ 1~mm and $\pm $ 3~mm offsets for $\vec{a_1}$ and $\vec{a_2}$. The cross check of the simulation is also done with COMSOL and shown for the optimum position and 1~mm offsets with open cross symbols.}
\label{fig9}
\end{figure}

Figure~\ref{fig9} illustrates the ratio of the elastic mirror motion displacement to the displacement in the case of rigid body motion as a function of frequency. The plot includes data for optimally positioned beams on the KAGRA test mass, as well as offsets of  $\pm $1 mm and  $\pm $3 mm from the optimal positions. For this study, we chose the symmetric position from the center, as asymmetrical positions of the beam spots can induce rotation effects due to torque. In this paper, we focused on the symmetry case to minimize the rotation effect. The rotation effect was estimated using an analytical equation, as shown in Eq.~\ref{eqrot}. The results obtained from ANSYS and COMSOL simulations were consistent. We assumed a symmetric offset from the optimal point in this study, as asymmetrical forces can cause torque and rotate the mirror. During the observation, we monitored the position of the beam spots using a telephoto camera. In the O3GK phase, the systematic errors arising from $S_\text{Ela}(f, \vec{a}, \vec{b})$ and $S_\text{Rot}(f, \vec{a}, \vec{b})$ were relatively negligible. However, in the O4 phase, these effects need to be included due to the improved calibration uncertainty.
Finally, the excitation signal generated by the photon calibrator, $x^\text{(Pcal)}$, was defined to measure the transfer function of the differential arm length. It is expressed as:
\begin{align}
x^\text{(Pcal)} = \frac{x_\text{Tx1} + x_\text{Tx2} + x_\text{Rx1} + x_\text{Rx2}}{2},
\label{eq:xpc}
\end{align}
where each term is defined in Eq.~\ref{eqx}. The estimated noise budget for the photon calibrator is shown in Table ~\ref{pcal_nb}. The estimated total systematic error arising from the calibration instruments is 2.0\%.
\begin{table}[h!]
\centering
\caption{Noise budget of each component. We calculated the noise propagation with Eq.\ref{eq:xpc}. Main errors are caused by Efficiency matrix}
\label{pcal_nb}
\begin{tabular}{|c|c|c|}
\hline
Mirror mass & $M$  & 0.041\%\\
\hline
Incident angle & $\theta$ & 0.00059\%\\
\hline
Gold standard response & $\rho_{\rm GS}$ & 0.30\%\\
\hline
Response ratio with working standard & $\alpha_{\text{WSK}}$ &1.0\% \\
\hline
Response ratio with University of Toyama system &  $\alpha_{\text{Toyama}}$  & 0.01\% \\
\hline
Efficiency matrix &$\epsilon^{(1)}_{\rm Tx}$&1.0\% \\
 &$\epsilon^{(2)}_{\rm Tx}$ & 0.6\%\\
 &$\epsilon^{(1)}_{\rm Rx}$& 1.0\%\\
&$\epsilon^{(2)}_{\rm Rx}$ & 0.6\% \\
\hline
Total & $x^\text{(Pcal)}$ & 2.0\%\\
\hline
\end{tabular}
\end{table}
\section{Discussion}
The development of a new photon calibrator with increased maximum power output to 20W opens up possibilities for new applications and technologies. In previous studies, LIGO and Virgo utilized photon calibrators with laser powers of 2W and 3W, respectively. With the increased power, the new photon calibrator offers advantages such as improved SNR for response function measurements, monitoring of time-dependent interferometer response, and hardware injection tests.

For response function measurements, the higher power laser allows for better characterization of the low-frequency and high-frequency regions with sufficient SNR. This leads to reduced uncertainties in these frequency ranges. Monitoring the time-dependent interferometer response involves injecting sine wave signals into the system and monitoring the resulting amplitudes. With the higher power photon calibrator, KAGRA can support a greater number of monitoring signals compared to previous systems, providing more comprehensive insight into the time-dependent behavior of the interferometer. In hardware injection tests, the photon calibrator can be used to verify the analysis pipeline by injecting simulated signals and confirming the accuracy of the extracted parameters.

Although the accuracy of beam positions is not currently a dominant systematic error in KAGRA's observations, it becomes more relevant in cryogenic studies. During the cooling process, the positions of the end test masses can shift due to thermal expansion effects in the suspensions. The expected displacement from thermal expansion is approximately 5mm. To mitigate the impact of elastic deformation, it is necessary to adjust the beam positions during the cooling process. This aspect will be further investigated and demonstrated in future experiments.

Regarding the systematic errors of the time-dependent interferometer model, they contribute significantly to KAGRA's overall error budget, accounting for 20\% and 20 degrees of the major sources of error. Even if we assume a systematic error of 3\% from the photon calibrator, its contribution to Observation 3 is not dominant. However, LIGO has achieved errors on the order of 1\% and 1 degree. To verify the response of the system, they employ two methods. One method involves crosschecking with a Newtonian Calibrator, which serves as an independent calibration source. The other method involves crosschecking with independent power standard institutes. Standard detectors are circulated among the standard institutes of different countries to ensure the consistency of laser powers through independent power standardization processes.

\section{conclusion}
A detailed understanding of waveform accuracy and the response of gravitational wave detectors is crucial for studying the physics of gravitational waves and exploring astronomical and cosmological phenomena. The photon calibrator method has emerged as a modern and accurate approach to calibration. In order to further improve the calibration accuracy, KAGRA has implemented several new techniques: (i)A 20W continuous-wave laser has been adopted to enhance the signal-to-noise ratio of high-frequency responses. This upgrade allows for more precise measurements in the high-frequency regime, thereby reducing uncertainties in the calibration process. (ii)Independent control of the upper and lower beams has been implemented to verify the response of the rotation component. By separately controlling these beams, KAGRA can validate and understand the rotation response of the interferometer, leading to a more accurate model of its behavior. (iii)A beam position monitoring and controlling system, utilizing quadrant photodiodes and pico-motors, has been employed. This system enables precise monitoring and control of the beam positions, contributing to overall calibration accuracy.
These advanced techniques have provided KAGRA with a better understanding of the accurate model for waveform calibration. In this paper, the relative power noise and harmonic noises were characterized, and the measured results satisfied the requirements for worldwide Observation 3.

During the observation phase, KAGRA exchanged the laser power standard detector with those from LIGO and VIRGO. This exchange facilitated the calibration of the relative amplitudes of each interferometer, promoting consistency and accuracy across different gravitational wave observatories.

Finally, the total noise contribution from the photon calibrator was estimated to be 2.0\%. By achieving a reduction in the systematic error from the instruments that is lower than the total systematic error in O3GK, KAGRA has demonstrated significant progress in minimizing the overall systematic uncertainties.

\begin{acknowledgments}
We would like to express our gratitude to Richard Savage, Evan Goetz, Jeff Kissel, and Peter King in LIGO for supporting us in developing a design for the KAGRA photon calibrator. We also would like to express Ayako Node, Ayako Ueda, Iwao Murakami, and Hirokazu Murakami to support this study. This work was supported by MEXT, JSPS Leading-edge Research Infrastructure Program, JSPS Grant-in-Aid for Specially Promoted Research 26000005, JSPS Grant-in-Aid for Scientific Research in Innovative Areas 2905: JP17H06358, JP17H06361, and JP17H06364, JSPS Core-to-Core Program A. Advanced Research Networks, JSPS Grant-in-Aid for Scientific Research (S) 17H06133, the Mitsubishi Foundation, the joint research program of the Institute for Cosmic Ray Research, University of Tokyo, National Research Foundation (NRF) and Computing Infrastructure Project of KISTI-GSDC in Korea, Academia Sinica (AS), AS Grid Center (ASGC), and the Ministry of Science and Technology (MoST) in Taiwan under the following grants: AS-CDA-105-M06, 110-2636-M-008-001, 110-2123-M-007-002, the KAGRA collaboration, the LIGO project, and the Virgo project.
\end{acknowledgments}

\bibliography{report}

\end{document}